\documentclass[prl,twocolumn,showpacs,preprintnumbers,amsmath,amssymb]{revtex4}
\usepackage[dvips]{graphics}

\begin{document}

\title{Time, (Inverse) Temperature and Cosmological Inflation as Entanglement}

\author{Vlatko Vedral}
\email{vlatko.vedral@gmail.com} 
\affiliation{Clarendon Laboratory, University of Oxford, Parks Road, Oxford OX1 3PU, United Kingdom and\\Centre for Quantum Technologies, National University of Singapore, 3 Science Drive 2, Singapore 117543 and\\
Department of Physics, National University of Singapore, 2 Science Drive 3, Singapore 117542}

\date{\today}

\begin{abstract}
We present arguments to the effect that time and temperature can be viewed as a form of quantum entanglement. Furthermore, if temperature is thought of as arising from the quantum mechanical tunneling probability this then offers us a way of  dynamically``converting" time into temperature based on the entanglement between the transmitted and reflected modes. We then show how similar entanglement-based logic can be applied to the dynamics of cosmological inflation and discuss the possibility of having observable effects of the early gravitational entanglement at the level of the universe.   
\end{abstract}

\pacs{03.67.Mn, 03.65.Ud}% PACS, the Physics and Astronomy

\maketitle                           %display desi d

\section{Introduction: the Church of Higher Hilbert Space} 

There is now an extensive amount of literature on entanglement in many-body systems \cite{Amico}. We have a good understanding of both how to quantify as well as qualify entanglement in complex systems.
The universe is of course the best example we have of a large complex many-body system and many of the techniques developed for quantifying and scaling of entanglement have also been applied to it (see for instance \cite{Diaz} and references therein). 
It is sometimes hard to see any connections between disparate results in the fields of quantum information and cosmology, which is why it might be beneficial to every once in a while take a broader 
perspective and summarize some aspects of our understanding. Here I would like to ask: can what we sometimes think of as different (cosmologically) relevant physical parameters actually be thought of as just different 
instances of quantum entanglement? 

In what follows I would like to recount the arguments that time and temperature can indeed be thought of as forms of entanglement. This is exciting for two reasons. One is that these potentially completely different entities can be seen 
to have the same common origin (in entanglement). It is always pleasing to be able to postulate no more phenomena than one needs to account for all observations (Occam). Secondly, however, claiming that
entanglement is at the root of these quantities might lead us to some observable consequences especially and most excitingly at the cosmological level. We will explore this in the second half of this paper. Finally, we outline how the fluctuations in the Cosmic Microwave Background (CMB) radiation can be used to witness entanglement at the cosmological level. The following two sections are largely a review of the existing material, though mainly from the author's own perspective. The last three sections present new material, by first unifying the arguments of the proceeding two sections and then extending them to cosmology and witnessing entanglement.  

First of all, I would like to set the scene by explaining the picture that is affectionately referred to (by the quantum information community) as the Church of Higher Hilbert Space. This picture is the expression of the fact that any mixed state (here written in its eigen-expansion)
\begin{equation}
\rho_1 = \sum_n r_n |r_n\rangle\langle r_n| 
\end{equation}
can (at least in principle) be represented as a reduction from a pure state existing on a Higher Hilbert space
\begin{equation}
\rho_1 = tr   |\Psi\rangle\langle \Psi|_{12} 
\end{equation}
where
\begin{equation}
|\Psi\rangle_{12} = \sum_n \sqrt{r_n} |r_n\rangle_1 \otimes |\phi_n\rangle_2 \; .
\end{equation}
The entropy of $\rho_1$, $S(\rho_1) = -\sum  r_n \ln  r_n$, quantifies the entanglement between the system (labeled by index $1$) and the extension (which itself is non-unique and is labeled by index $2$).  The reason for using the word "Church" is that, though the statement that ``we can always write a mixed state as a reduction of a pure one" looks like a tautology (and hence always true), it is actually an expression of our belief that this extension to purity could always be performed in practice (this, of course, is an open question since we might run out of resources to perform the required purification). 

The Hilbert Space extension is an important mathematical technique when proving many results in quantum information theory, ranging from calculating entanglement
measures to proving security of quantum cryptography and establishing various quantum channel capacities. But we now proceed to show how it can also be used 
to argue that two concepts we think of as fundamental, time and temperature, can also be seen as instances of entanglement with extended Hilbert space.

\section{Time as Entanglement} 

The method of viewing time as entanglement simply encapsulates the fact that we never observe time directly. We usually observe the position (of the hand of the clock, the sun or the stars) or some other observable of a periodically evolving 
system. Therefore, when we are timing the evolution of the system under consideration, we are always talking about the system's states with respect to the state of the clock. The clock in this case will provide the extending Hilbert space within
which nothing ever evolves. However, as we show below, the state of the system will evolve relative to the state of the clock. Here we follow the work of Page and Wootters \cite{Wootters}, although essentially the same logic is built into arguments of Banks \cite{Banks} and Brout \cite{Brout} (for a pedagogical review see \cite{Briggs}). The germs of this idea go back to a paper by Mott \cite{Mott}, where he used the time independent Schr\"odinger equation to derive trajectories of alpha particles in a cloud chamber (the point being that the background atoms in the chamber act as a clock recording the position and hence the time of the passing alpha particle).  

Suppose therefore that we are in an eigenstate $|\Psi_{sc}\rangle$, of a Hamiltonian, $H$, consisting of two different subsystems, call them the system ($s$) and the clock ($c$). Suppose further that the 
interaction between the system and the clock is negligible so that $H=H_s + H_c$ (this is what in fact defines a good clock, namely that it is, at least to a high degree, independent of the system). We assume without any loss of generality that  $H|\Psi_{cs}\rangle = 0$ (all this does it set the overall phase which is in any case an unobservable quantity). 

Imagine furthermore that the state $|\Psi_{sc}\rangle$ has a special suitably chosen form:
\begin{equation}
|\Psi_{cs}\rangle = \sum_{\tau} |\psi_s (\tau) \rangle \otimes |\psi_c (\tau)\rangle \; .
\end{equation}
The normalization $N$ is absorbed in the states themselves so that $\langle \psi_c (\tau) |\psi_c (\tau)\rangle = \sqrt{N}$. Here, and without any loss of generality, we are using $\tau$ as a discrete parameter (though it can also always be made
continuous).  

We can now postulate that the clock states above are constructed in such a way that 
\begin{equation}
e^{-iH_c} |\psi_c (\tau)\rangle = |\psi_c (\tau+d\tau)\rangle 
\end{equation}
i.e. the clock Hamiltonian generates shifts between one clock time and the immediate next clock time (note that this is just a mathematical property of the states with respect to the Hamiltonian, there is actually no real temporal evolution taking place yet). 
An obvious clock to choose is a quantized rigid rotor, but our discussion is completely generic and does not require us to confine ourselves to 
anything that resembles the traditional classical clock. 

Now we look at the evolution of the system relative to the states of the clock (the relative state of the system in the same Everett sense \cite{Everett}):
\begin{eqnarray}
i\hbar \frac{d}{d\tau} | \psi_{s}\rangle & = & i\hbar \frac{d}{d\tau} \langle \psi_c| \Psi_{sc}\rangle\\
 & = & - \langle \psi_c| H_c| \Psi_{sc}\rangle \\
 & = & \langle \psi_c| H_s - H | \Psi_{sc}\rangle \\
 & = & \langle \psi_c| H_r \Psi_{sc}\rangle \\
 & = & H_s| \psi_{s}\rangle
\end{eqnarray}
and so the system undergoes the Schr\"odinger type evolution relative to the ticking of the clock. Time therefore arises internally without the need for any global time. This kind of argument is therefore potentially important in cosmology where there are presumably no clocks to measure time outside of the universe. The cosmological time itself then has to emerge from within, as in the calculation above. 

An important subtlety is that the clock need not encompass the rest of the universe, though it can include it if required (as in \cite{Brout}). This means that the above argument would work even if the state of the system and the clock was mixed when the rest of the universe was traced out. All that matters is the relative state of the system with respect to the clock.  Next we show how temperature can likewise arise without the need for having an overall temperature. 

\section{Temperature as Entanglement} 

Obtaining temperature from a temperature-less universe requires us again to start from a pure state of two subsystem and then look at the subsystems individually. First of all, it is clear that a thermal state at temperature $T$ that is described as
\begin{equation}
\rho_T = \sum_n p_n |E_n\rangle\langle E_n|
\end{equation}
where $p_n = e^{-\beta E_n}/Z$ ($\beta = 1/k_B T$), can always be obtain from an extension of the form
\begin{equation}
|\Psi\rangle = \sum_n \sqrt{p_n} |E_n\rangle \otimes |\phi_n\rangle \; .
\end{equation}
From what we said before it follows that temperature $T$ and entanglement (as measured by the entropy of the reduced states) are directly related: the higher the temperature, the higher the entanglement between the two subsystems. 

This simple argument can, in fact, be made to resemble the ``timeless time" argument even further. The bonus will be that the Gibbs-Boltzmann distribution will arise naturally providing we make a few assumptions (to be detailed in what follows).  

Imagine we divide the total universe into a small system ($s$) and a large rest ($r$). The attributes ``small" and ``large" will be quantified below. 
Let us again assume that the interaction between the system and the rest is small enough to be negligible and that the total state is a zero energy eigenstate
$(H_s + H_r)|\Psi_{sr}\rangle = 0$. The reason for this will become transparent shortly (we recall that in the clock argument this was needed because a good clock neither affects nor is affected by
the evolution of the system - at least to within a good approximation). 

Now construct $|\Psi_{sr}\rangle$ as a superposition of energy eigenstates of the system $|E_n\rangle$ correlated to the states of the rest with energy $-E_n$ (since 
the sum has to add up to zero - here is where we need the assumption that the interaction Hamiltonian between the two vanishes). The total state can be written as
\begin{equation}
|\Psi_{sr}\rangle = \sum_n |E_n\rangle \otimes \sum_{m=1}^{D(-E_n)} |\psi_{nm}\rangle \; ,
\end{equation}
where the states of the system are not normalized so that $\langle E_m|E_n\rangle = N \delta_{nm}$. 
The index $m$ for the rest takes into account the fact that the rest is huge compared with the system and there may be many degenerate states whose energy is $-E_n$. The
degree of degeneracy will be labeled as $D(-E_n)$. To obtain the state of the system, $\rho_s$ we trace out the rest, i.e.
\begin{eqnarray}
\rho_s & = & \sum_n |E_n\rangle\langle E_n|  \sum_m tr (|\psi_{nm}\rangle\langle\psi_{nm}|)\\
& = &   \sum_{nm} \langle\psi_{nm}|\psi_{nm}\rangle |E_n\rangle\langle E_n|\\
& = &    \sum_n D(-E_n)|E_n\rangle\langle E_n| \; .
\end{eqnarray}
We now assume that the energies $E_n$ are small enough that we can expand to the first approximation (this is one of the two central assumptions leading to Gibbs-Boltzmann weights as we will shortly see):
\begin{eqnarray}
D(-E_n) & = & D(0) - \frac{dD}{dE_n}E_n = D(0) \left(1 - \frac{1}{D(0)}\frac{dD}{dE_n}E_n \right) \nonumber\\
& \approx & D(0) \exp\left\{- \frac{1}{D(0)}\frac{dD}{dE_n}E_n\right \}   \; .
\end{eqnarray} 
The second central assumption is that the function whose first order expansion is $f(x) = 1 - x +...$ is in fact the exponential $e^{-x}$ (there are of course infinitely many functions that have the same first order Taylor expansion;the exponential can be further justified by reuqiring that $f(x+y) = f(x)f(y)$, namely that densities of independent systems get multiplied).   
We can now define 
\begin{equation}
\beta = \frac{1}{D(0)}\frac{dD}{dE_n} \; ,
\end{equation}
which is our effective inverse temperature. We can rewrite this in an even more transparent way as
\begin{equation}
\beta = \frac{\ln D(0)}{dE_n}
\end{equation}
where we now have the standard statistical definition of inverse temperature as the derivative of entropy with respect to energy. The state of the system now emerges to be
\begin{eqnarray}
\rho_s & = & \sum_n \frac{e^{-\beta E_n}}{Z}|E_n\rangle\langle E_n|
\end{eqnarray}
where $Z=\sum_n e^{-\beta E_n}$ is the partition function which arises from the normalization $N$. Just like we noted in the case of time, there is here no need to start from an entangled state of the system and the rest; a mixture will suffice just as well for the above argument \cite{Dahlsten}. However, one can always assume
that the state is purified to include everything in the universe, so that the rest is indeed the rest of the universe excluding the system. We will revisit this argument when we discuss cosmological inflation. 

We have now seen that both time and temperature can arise from entanglements between the system under consideration and another suitably chosen system. 
But could the two (time and temperature) be related more directly? Namely, is there a physical process that can convert time into temperature (or vice versa)? An interesting possibility is to view a system that dynamically
tunnels through a potential barrier. The state of the system is a superposition of the transmitted and the reflected wave. However, suppose that we only have access to the transmitted wave. Then, we actually need
to trace out the reflected wave in which case the transmitted state is a mixed one (and can therefore be thought of as being at some finite temperature). 

The presence of entanglement in this example is a bit more subtle. It can be seen to arise from the second quantized notation of the tunneling particle having an amplitude to tunnel (i.e. to be transmitted through the barrier) and another amplitude not to tunnel (i.e. to be reflected by the barrier).
The state can then be written as
\begin{equation}
\sqrt{r} |1\rangle_r |0\rangle_t + \sqrt{t} |0\rangle_r |1\rangle_t
\end{equation}
where the $r$ and $t$ subscripts indicate the reflected and transmitted modes respectively. When we trace out the reflected mode, we obtain a mixed state of the transmitted mode. It is interesting that this process
of conversion of time into temperature by quantum tunneling was recently employed by Parikh and Wilczek \cite{Wilczek} to explain the Hawking radiation \cite{Hawking} and the resulting temperature of a black hole.

\section{Conversion of time into temperature by tunneling}

We now briefly summarize the argument by Parikh and Wilczek \cite{Wilczek}. We imagine that a particle-anti-particle pair was created inside the black hole, close to the event horizon, and that the particle is then able to tunnel out.
We proceed to calculate the probability for this to happen. The inverse of this process is the creation of the pair outside of the event horizon and that the anti-particle tunnels into the black hole. The two processes 
will have the same probability (since their amplitudes are presumably complex conjugates of one another). We now proceed to explain how this is calculated. 

The main ingredient is the formula for quantum tunneling. The reader will recall that, in the WKB approximation, the trial solution to the Schr\"odinger equation
\begin{eqnarray}
\frac{d^2 \psi (x)}{dx^2} + k^2(x) \psi(x) = 0 
\end{eqnarray}
where $k^2 = -2mV(x)/\hbar^2$ (we assume $E=0$)
is given by
\begin{eqnarray}
\psi (x) = \psi_0 \exp\left\{ i\int_0^x k(x')dx' \right\} \; .
\end{eqnarray}
This assumes that $|k'|<<k^2$ (i.e. small de Broglie wavelength). The tunneling rate is, within this approximation, defined as the ratio of the outgoing to the incoming intensity of particles and this is given by
\begin{equation}
\Gamma = \exp\left\{-2 \int_{r_{in}}^{r_{out}} k(r)dr\right\} \; .
\end{equation}
Here $r_{in}$ and $r_{out}$ are the boundaries of the potential. It is now clear that since this is an exponential dependence of the probability on the wavenumber, we can equate it to the Boltzmann
weight ($\exp -E/k_BT$) and thereby obtain an ``effective" temperature. This is the gist of the  Parikh-Wilczek argument.

To illustrate how to obtain the Hawking temperature we will now apply this formula to the scenario where a particle (antiparticle) tunnels out of (into) a black hole of mass $M$. In the former case, $r_{in}=2M$ and $r_{out}=2(M-\omega)$,  are the Schwarzshild radius before and after the particle (of frequency $\omega = m$, where we have set $\hbar=c=k_B=G=1$; we will reintroduce the constants shortly below) has tunneled out respectively. We will assume an otherwise flat potential. Evaluating the integral:
\begin{eqnarray}
\int_{r_{in}}^{r_{out}} k(r)dr & = & -Im \int_{r_{in}}^{r_{out}} \int_0^{k'} dk'(r)dr \nonumber\\
& = & -Im \int_{r_{in}}^{r_{out}} \int_0^{\omega} d\omega' \frac{dr}{1-\sqrt{2(M-\omega')/r}} \nonumber \; .
\end{eqnarray}
Here we have used the fact that $d H/d k = \dot r$, that $dH = -d\omega$ and that, finally, $\dot r = 1-\sqrt{2(M-\omega)/r}$. 

This integral can be solved (by using e.g. calculus of residues)  to yield:
\begin{equation}
\Gamma = \exp\{-8\pi \omega (M-\omega/2)\} \; .
\end{equation}
Equating this to the thermal distribution $\exp \{ \hbar \omega /kT\}$, and ignoring the $\omega^2$ contribution, gives us the following temperature (with all the relevant constants now in place):
\begin{equation}
T \approx \frac{\hbar c^3}{8\pi G k_B M} \; ,
\end{equation}
which is indeed the Hawking temperature for a black hole. 

There are two comments to make regarding the connection between tunneling and temperature. One is that the outgoing distribution is not quite thermal (since it contains a correction on the order of  $\omega^2$ that we ignored in last step). 
This correction is at least in principle observable. And if not for an actual black hole (for which even the Hawking temperature is tiny), this might be possible in one of the many analogue systems mimicking the black hole dynamics. 

Secondly, however, we have a more interesting possibility, namely that we can apply the same idea of particles tunneling but this time beyond the edge of the observable universe.
Could it be that this kind of process can also account for entropy of the observable universe?

\section{Cosmological Inflation} 

In quantum cosmology, we treat the universe as a quantum system, but then use the resulting model to compute some macroscopically observable parameter, such as the temperature of the Universe, or its density fluctuations (ultimately also measure by the temperature fluctuations in the CMB). If the universe is a closed system, its state ought to be pure, and it is then impossible to assign a (non-zero) temperature to it. However, if we imagine that the pure state belongs to the universe as a whole and we only observe a small section of the whole universe (as the theory of inflation might suggest) then it is clear how the observable universe could be in a mixed state to which it is then appropriate to assign a temperature. The suggestion to treat the observable universe as an open quantum system comes from Prigonine \cite{Prigogine}, though the line of argument we use here will be entirely different. 

The plan is to apply the temperature-as-the-result-of-tunneling argument to the universe.  The evolution of the universe follows two equations that are derived from Einstein's field equations assuming the cosmological principle, which tells us that the universe is 
homogeneous and isotropic (i.e. the same for all observers, which in turn fixes the metric to be used).

The two equations describing the evolution of the universe are known as the Friedmann equations: 
\begin{eqnarray}
3\left(\frac{\dot a}{a}\right)^2 + \frac{3kc^2}{a^2}  & = & \frac{8\pi G \rho}{c^2}\\
-2\frac{\ddot a}{a} - \left(\frac{\dot a}{a}\right)^2 - \frac{kc^2}{a^2} & = &  \frac{8\pi G p}{c^2}
\end{eqnarray}
Here $a$ is the scale-factor of the universe, $k$ its curvature (not to be confused with the Boltzmann constant $k_B$), $G$ Newton's gravitational constant, $c$ the speed of light, $\rho$ is the density of the universe and $p$ is the pressure. These are two equations with three unknowns (the pressure, the density and the scale factor). Usually we assume an equation of state relating the pressure and volume and then solve for the scale factor. Here however we will follow a different route.  

We now present an argument for the temperature of the universe that mirrors the Parikh-Wilczek black hole calculation. Imagine that the temperature is a consequence of matter tunneling into the observable universe (this argument was originally used to calculate the probability for the universe to tunnel into its own existence \cite{Vilenkin,Tryon}). This process can be described by a quantized version of the first of the Friedmann equations (the quantum version is known as the Wheeler-DeWitt equation) \cite{Atkatz}:
\begin{equation}
\left\{ \frac{\partial^2}{\partial a^2} -\left(\frac{3\pi}{2G}\right)^2 a^2\left(1-\frac{a^2}{a^2_0}\right)\right\}|\psi (a)\rangle = 0 \; , 
\end{equation}
where $a_0^2 = G$ (assumed for simplicity) is the size of the tunneling barrier (as is customary we have again set $c=1$).  The solution of the above equation is known as the wave-function of the universe \cite{Hartle}. 
The probability to tunnel through the potential $(\frac{3\pi}{2G})^2 a^2(1-\frac{a^2}{a^2_0})$ is given by
\begin{equation}
p = \exp \left\{-\frac{3\pi}{2G} \int_0^{a_0} da  a \sqrt{1-\frac{a^2}{a^2_0}}\right\} \; ,
\end{equation}
which leads us to the tunneling probability in the universe   
\begin{equation}
p\approx e^{-\frac{c^5}{\hbar G^2 H}} \; ,
\end{equation}
where $H=\dot a/a$ is the Hubble parameter.  Note that this is usually applied to the beginning of the universe with some fixed initial value of $H$. In our case, this formula holds at all times and describes the tunneling between the observable universe and the rest.  Writing this in the exponential Gibbs-Boltzmann form 
\begin{equation}
p\approx e^{-\frac{M_uc^2}{k_BT_u}}\; 
\end{equation}
where $M_u$ is the mass of the observable universe and $T_u$ its temperature. Using the fact that $M_u = c^3/4GH$ \cite{Valev} we obtain
\begin{equation}
T_u = \frac{\hbar H}{4\pi k_B} \; .
\end{equation}
As already noted, this temperature is time dependent (as the result of the time dependence of $H$). 
We briefly point out that the mass of the universe can be arrived at by different methods to be about $10^{53}kg$ (see e.g. \cite{Valev}) which is in a pretty good agreement with the formula used here (and which can almost be obtained by dimensional analysis by combining $c$, $G$ and $H$ into a quantity with dimensions of mass). 

We will now use this temperature and assume the universe to be a black body (here we follow the argument given in \cite{Modak}). This will then be inserted into the continuity equation (which is basically the First law of thermodynamics and is derivable from the Friedmann equations) which is of the form:
\begin{equation}
\frac{dQ}{dt} = \frac{d}{dt}(\rho V) + p \frac{dV}{dt}
\end{equation}
where $V$, $p$ and $\rho$ are the volume, pressure and density of the universe respectively. If the universe is truly an isolated system then $dQ=0$ and the left hand side of above would vanish (which is what is normally assumed). 
However, if we think of just the observable universe then the theory that best fits the current observation is based on the idea of inflation. Namely, at the very beginning the universe was supposed to have undergone a rapid expansion which stretched the space-time fabric faster than the speed of light. As a consequence what we call the observable universe is only a part of the total universe, the rest of it being outside of our light cone. If we suppose that the universe has always been quantum mechanical, then the observable universe should be entangled with the rest.  

Furthermore and in line with the Church of Higher Hilbert Space picture, all the mixedness (entropy) in our observable universe comes from tracing out the rest.
This would then provide us with the term  $dQ/dt$ (see also the discussion in \cite{Prigogine}). The rest in this case is the component of the universe that lies outside the horizon, and we assume that the universe is at a temperature derived from the tunneling argument above. The change in time comes from the fact that the universe is evolving, which in turn affects the entanglement between the observable universe and the rest (and therefore leads to a changing temperature). 
Since we are assuming that $Q$ has the black body spectrum, we can then use the Stefan-Boltzmann law to write the rate of change of heat as
\begin{equation}
\frac{dQ}{dt} = A \sigma T^4
\end{equation}
where $A$ is the area of the observable universe, $T=\hbar H/2\pi k_B$ is the related Hawking temperature and $\sigma$ is the Stefan constant.   Here the area will be written as $A=4\pi r^2_u$ where the radius of the observable universe will be taken to be
$r_u = c/H$. We can also express the volume of the universe as $V=4/3\pi r^3_u$. 

The first law now reads 
\begin{equation}
\frac{3\sigma}{c} \left(\frac{\hbar H}{4\pi k}\right)^4 H^5 = \frac{d}{dt}(\rho) + 3 (\rho +p) \frac{\dot a}{a}
\end{equation}
Using the first of Friedmann equations (with the curvature term $k$ set to zero) and the fact that $H=\dot a/a$ we can transform the above to:
\begin{equation}
\frac{\dot \rho}{\rho} + 3(1+\omega) \frac{\dot a}{a} = 3\omega_c \frac{\dot a}{a} 
\end{equation}
where $p=\omega \rho$ is the equation of state relating the pressure and density of the universe, and $\omega_c = \hbar G^2/45c^7 \rho$ is the (time-dependent) critical density. We now see that the term due to $\frac{dQ}{dt}$ (which is on the right hand side of the equation) effectively acts as negative pressure (countering the second term on the left hand side). We can solve this equation for $\rho$ to obtain:
\begin{equation}
\rho = \frac{D a^{-3(1+\omega)}}{1+\frac{\alpha D}{1+\omega}a^{-3(1+\omega)}} 
\end{equation}
where $D$ is just a constant and $\alpha = \hbar G^2/45c^7 $ . If we assume the equation of state for ordinary radiation ($\omega = 1/3$) and that $\alpha D >> a^4$ we obtain that 
\begin{equation}
\rho = \frac{4}{3\alpha} 
\end{equation}
which is a constant density. This allows us to integrate the first Friedmann equation leading to an exponential expansion of the scale factor
\begin{equation}
a(t) \propto e^{Ht}
\end{equation}
where $H = \sqrt{32 \pi G/9 c^2\alpha} = 10^{45} s^{-1}$. This is the expansion rate required for the inflationary period. It is driven by the negative pressure provided by  $\frac{dQ}{dt}$. According to this model the inflation stops once this term becomes negligible, which is when $a^4 >> \alpha D$ and here we have the radiation dominated era. The open system evolution of the universe therefore naturally accounts for the inflationary expansion so long as we allow the heat exchange term to be based on the back body radiation due to the tunneling of stuff into the observable universe thereby leading to Hawking's temperature of the observable universe. 

Two warnings are appropriate here. Firstly, it is hard to trust the above semiclassical model at very early times of the evolution, when the full quantum gravitational effects might be significant. In the absence of the theory of quantum gravity, it might be more appropriate to use quantum field theory in curved spacetime (see for instance, \cite{Parker}), thought the main advantage of the treatment above is its simplicity and elegance. The last section will in fact reinforce the need for quantum gravity. Secondly, current astronomical observations \cite{Baumann} put a bound on the value of the inflationary Hubble parameter in certain inflationary models. This value is about ten orders of magnitude smaller than the above one of $H = 10^{45} s^{-1}$. However, the bounds come from evaluating the ratio of the scalar to tensor perturbations that are based on a scalar field driven inflation, a mechanism that is different to the one considered here (see also \cite{Modak-reply}). 

The idea that entanglement may be responsible for a number of fundamental parameters and processes is intriguing in its own right, and, as noted, it might help us reduce the number of mysteries that we have to explain. 
But, as scientists we should be asking if there are any observable effects of any of the above ideas. Obviously it is hard to see how to confirm the entanglement between the observable and non-observable universe (if there is such a thing). However, 
we might legitimately ask if entanglement could be witnessed indirectly - at least in principle - through its effect on the CMB profile. This is the topic we turn to next.

\section{Cosmological witnesses of entanglement}

The possible effects of quantum gravity on the CMB spectrum are very much  discussed and analyzed (see e.g. \cite{Woodward}). Here we follow the logic of constructing macroscopically observable entanglement witnesses that might be inferred from the CMB. 
We expect that the effects of quantum physics and gravity were important in the very early stages of the universe and were then possibly amplified by the process of inflation. Cosmologists in fact believe that all the structure in the universe come from
the original quantum fluctuations whose effect was then amplified by gravity. But how do we know that the correlations we observe are due to quantum correlations and not just of an entirely classical origin? (After all we said that both time and temperature can arise in the same way from a mixed, classically correlated state). Here we present a simple argument.   

Suppose that we are given a thermal state $\rho_T = p|\Psi_0\rangle\langle \Psi_0 | + (1-p) \rho_{rest}$, where $|\Psi_0\rangle$ is the ground state, $p=\exp(-E_0/k-BT)/Z$ is the usual Boltzmann weight and $\rho_{rest}$ involves all higher levels. A very simple entanglement witness can now be derived by noting that if
\begin{equation}
S(|\Psi_0\rangle ||\rho_T) < S(|\Psi_0\rangle ||\rho_{sep})=E(|\Psi_0\rangle)
\end{equation}
where $S(\sigma ||\rho)$ is the quantum relative entropy \cite{Vedral1}, then the state $\rho_T$ must be entangled (as it is closer to $|\Psi_0\rangle$ than the closest separable state, which we denoted as $\rho_{sep}$). $E(\rho)$ is the relative entropy of entanglement of $\rho$ \cite{Vedral1,Vedral2}. After a few simple steps, the above inequality leads to another inequality, satisfied by entangled thermal states $\rho_T$,
\begin{equation}
-\ln p < E(|\Psi_0\rangle) \; .
\end{equation}
Exploiting the fact that
\begin{eqnarray}
p & = & \frac{e^{-E_0/k_BT}}{Z} = e^{-(E_0+k_BT\ln Z)/kT} \geq e^{-(U+F)/k_BT} \nonumber\\
& = & e^{-S/k_B}\nonumber \; ,
\end{eqnarray}
where $F=-k_BT\ln Z$ is the free energy and $S=(F+U)/T$ is the entropy, we finally obtain the inequality
\begin{equation}
S(\rho_T) < k_B E(|\Psi_0\rangle)
\end{equation}
which, if satisfied, implies that $\rho_T$ is entangled. We now have a very simple criterion which can be expressed as follows: if the entropy of a thermal state is lower than the relative entropy of its ground state (multiplied by the Boltzmann constant $k$), then this thermal state contains some form of entanglement. 

Here we are not concerned with the type of entanglement we have (e.g. bi-partite or multipartite, distillable or bound), but we only what to confirm that the state is not fully separable. It is also very clear that if the ground state is not entangled, this witness will never detect any entanglement (since entropy is always a non-negative quantity), even though the state may in reality be entangled for some range of temperatures.

The entanglement witness based on entropy, though at first sight very simple, is nevertheless rather powerful as it allows us to talk very generally about temperatures below which we should start to detect entanglement in a very generic macroscopic system.
Since entropy is lower at low temperatures, this is the regime where we expect the witness to show entanglement. Let us look at the typical examples of ideal bosonic and fermionic gases. Non-ideal systems behave very similarly, with some for us unimportant corrections. At low $T$, the entropy scales as (see e.g. \cite{Landsberg})
\begin{equation}
S \sim N \left(\frac{k_BT}{\hbar \tilde \omega_{F,B}}\right)^{p_{F,B}}
\end{equation}
where $F,B$ refer to fermions and bosons respectively, $N$ is the (average) number of particles, $\tilde \omega$ is some characteristic frequency which is a function of the spectrum (its form depends on the details of the system) and $p\geq 1$. The fact that this form is the same for more general systems is due to what is known as the third law of thermodynamics (see \cite{Landsberg} for example) stating that the entropy has to go to zero with temperature. We now consider how entanglement scales in the ground state for fermions and bosons. If the number of particles is comparable to the number of modes, this typically means that $E\sim N$. The entropy witness then yields a very simple temperature below which entanglement exists for both fermions and bosons,
\begin{equation}
k_BT <  \hbar \tilde \omega_{F,B}
\label{ent-crit}
\end{equation}
This kind of temperature has been obtained in a multitude of different systems, ranging from spin chains, via harmonic chains and to (continuous) quantum fields. Its universality is now justified from a very simple behaviour of entropy at low temperatures.

It is however hard to observe the entropy of any object. A physically accessible information is more easily found in quantities such as the heat capacity \cite{Wiesniak},
\begin{equation}
C=T\frac{\partial S}{\partial T}
\end{equation}
In terms of the heat capacity, eq (\ref{ent-crit})  implies that the values of the heat capacity below
\begin{equation}
C< C_{crit} = \left(\frac{k_BT}{\hbar \tilde \omega_{F,B}}\right)^{p_{F,B}}
\end{equation}
cannot be accounted for without quantum entanglement. When it comes to cosmological observations, what is actually measured are the (relative) fluctuations in the temperature of the CMB: $\delta T/T$. 
The following relationship exists between the temperature fluctuations and the heat capacity: 
\begin{equation}
\frac{\Delta T}{T} = \sqrt{\frac{k_B}{C}}
\end{equation}
Which allows us to restate the criterion for entanglement as
\begin{equation}
\left(\frac{T}{\Delta T}\right)^2 \le  \left(\frac{k_B T}{\hbar \tilde  \omega_{F,B}}\right)^{p_{F,B}}
\end{equation}
We will choose $p_{F,B}=1$. According to current estimates $\delta T/T = 10^{-5}$. This in turn leads to $\tilde  \omega_{F,B} \le 10^2 Hz$ where we have used $T \approx 3K$ for the CMB temperature. Gravitational waves
are expected to have frequencies between between $10^{-15} Hz \le \omega \le 10^4 Hz$ \cite{Thorne}. It is therefore possible that the present temperature fluctuations do indicate that most gravitational waves 
were entangled and therefore need to be treated intrinsically quantum mechanically (i.e classical stochastic treatment may not be sufficient). This is in agreement with a different kind of qualitative estimate which
also argues that quantum gravitational effects should not be negligible in the fluctuations of the CMB \cite{Krauss}. The Krauss-Wilczek argument however has to do with the effect of gravitons on the polarization of light in the CMB. The argument 
here, on the other hand, would need to rely on a resonant conversion of gravitons into photons, which then ``inherit" the entanglement from the quantum gravitational field and this resulting photonic entanglement is what is 
then measured through the fluctuations in the CMB radiation. How this might happen is well beyond the scope of the current work.

\textit{Acknowledgments}: The author acknowledges funding from the
National Research Foundation (Singapore), the Ministry
of Education (Singapore), the EPSRC (UK), the Templeton
Foundation, the Leverhulme Trust, the Oxford
Martin School, the Oxford Fell Fund and the European Union (the EU Collaborative Project TherMiQ, Grant Agreement 
618074). 
.


\begin{thebibliography}{1}
%
\bibitem{Amico} L. Amico, R. Fazio, A. Osterloh and V. Vedral, Rev. Mod. Phys. {\bf 80}, 1 (2008).
%
\bibitem{Diaz} P. F. Gonzalez-Diaz, C. L. Sig\"uenza and J. Martin-Carion, Phys. Rev. D {\bf 86}, 027501 (2012).  
%
\bibitem{Wootters} D. Page and W. Wootters, Phys. Rev. D {\bf 27}, 2885 (1983).
%
\bibitem{Banks} T. Banks, Nucl. Phys. B 249, 332 (1985).
%
\bibitem{Brout} R. Brout, Found. Phys. 17, 603 (1987); R. Brout, G. Horwitz,
and D. Weil, Phys. Lett. B 192, 318 (1987); R. Brout, Z. Phys.
B {\bf 68}, 339 (1987).
%
\bibitem{Briggs} J. S. Briggs and J. M. Rost, Found. Phys. {\bf 31}, 693 (2001). 
%
\bibitem{Mott} N. Mott, Proc. Roy. Soc.  A {\bf 126}, 79 (1929).
%
\bibitem{Everett} H. Everett, {\em On the Foundations of Quantum Mechanics}, (Ph.D. thesis, Princeton University, Department of Physics, 1957).
%
\bibitem{Dahlsten} Oscar C. O. Dahlsten, Cosmo Lupo, Stefano Mancini, Alessio Serafini, Entanglement typicality, arXiv:1404.1444. 
%
\bibitem{Wilczek}  M. K. Parikh and F. Wilczek, Phys. Rev. Lett. {\bf 85}, 5042 (2000). 
%
\bibitem{Hawking} S. W. Hawking,  Comm. Math. Phys. {\bf 43}, 199 (1975). 
%
\bibitem{Prigogine} I. Prigogine, Int. J. Theor. Phys. {\bf 28}, 927 (1989). 
%
\bibitem{Vilenkin} A. Vilenkin, Physics Letters B {\bf 117}, 25 (1982). 
%
\bibitem{Tryon} E. P. Tryon, Nature {\bf 246}, 396 (1973).
%
\bibitem{Atkatz} D. Atkatz, Am. J. Phys. {\bf 62}, 19 (1994). 
%
\bibitem{Hartle} J. B. Hartle and S. W. Hawking, Phys. Rev. D {\bf 28}, 2960 (1983). 
%
\bibitem{Valev} D. Valev, Estimation of the total mass and energy of the unvierse, arXiv:1004.1035v1. 
%
\bibitem{Modak} S. K. Modak and D. Singleton, Int. J. Mod. Phys. D {\bf 21}, 1242020 (2012); S. K. Modak and D. Singleton, Phys. Rev. D {\bf 86}, 123515 (2012).
%
\bibitem{Parker} L. E. Parker and D. J. Toms, {\em Quantum Field Theory in Curved Spacetime}, (Cambridge University Press, 2009). 
%
\bibitem{Baumann} D. Baumann et al. (CMBPol Study Team), AIP Conf. Proc.
{\bf 1141}, 10 (2009).
%
\bibitem{Modak-reply} S. K. Modak and D. Singleton, Phys. Rev. D {\bf 89}, 068302 (2014).
%
\bibitem{Woodward} R. P. Woodward, Rep. Prog. Phys. {\bf 72} 126002 (2009). 
%
\bibitem{Vedral1} V. Vedral, M. B. Plenio, M. Rippin and P. L. Knight, Phys. Rev. Lett. {\bf 78}, 2275 (1997).
%
\bibitem{Vedral2} V. Vedral and M. B. Plenio, Phys. Rev. A {\bf 57}, 1619 (1998).
%
%\bibitem{Landau} L. Landau and E. Lifschitz, {\em Statistical Physics Parts I and II}, (Pergamon Press, Oxford).
%
\bibitem{Landsberg} P. T. Landsberg, {\em Thermodynamics and Statistical mechanics}, (Dover, New York, 1990).
%
\bibitem{Wiesniak} M. Wiesniak, V. Vedral, and C. Brukner, Phys. Rev. B 78, 064108 (2008).  
%
\bibitem{Thorne} K. S. Thorne, {\em Gravitational Waves}, (Cornell University Library, 1995).
%
\bibitem{Krauss} L. M. Krauss and F. Wilczek, Phys. Rev. D {\bf 89}, 047501 (2014). 



\end{thebibliography}
\end{document}